\documentclass[12pt]{article}
\usepackage{epsfig}

\usepackage{graphicx}
\usepackage{cancel}
\usepackage{amssymb,amsmath}

\def\harr#1#2{\smash{\mathop{\hbox to .3in{\rightarrowfill}}
 \limits^{\scriptstyle#1}_{\scriptstyle#2}}}

\def\s2{\frac{1}{\sqrt2}}

\def\be{\begin{equation}}
\def\ee{\end{equation}}
\def\beqa{\begin{eqnarray}}
\def\eeqa{\end{eqnarray}}

\def\Tr{{\rm Tr \,}}

\def\Dsl{\,\raise.15ex\hbox{/}\mkern-13.5mu D} 
\def\d3{d^3}

\def\IR{\mathbb{R}}
\def\IC{\mathbb{C}}

\begin{document}

\vspace{.5cm}
\begin{center}
\Large{\bf Towards Noncommutative Linking Numbers Via the Seiberg-Witten Map}\\
\vspace{1cm}

H. Garc\'{\i}a-Compe\'an$^a$\footnote{e-mail address: {\tt compean@fis.cinvestav.mx}}, O. Obreg\'on$^b$\footnote{e-mail
address: {\tt octavio@fisica.ugto.mx}}, R.
Santos-Silva$^b$\footnote{e-mail address: {\tt
rsantos@fisica.ugto.mx}}
\\
[2mm]{\small \em $^a$Departamento de F\'{\i}sica, Centro de
Investigaci\'on y de Estudios Avanzados del IPN}\\
{\small\em P.O. Box 14-740, CP. 07000, M\'exico D.F., M\'exico.}
\\[4mm]
{\small \em $^b$Departamento de F\'isica, DCI, Universidad de Guanajuato\\
 C.P. 37150, Le\'on, Guanajuato, M\'exico.}\\

\vspace*{2cm}
\small{\bf Abstract} \\
\end{center}
In the present work some geometric and topological implications of
noncommutative Wilson loops are explored via the Seiberg-Witten map.
In the abelian Chern-Simons theory on a three dimensional manifold,
it is shown that the effect of noncommutativity is the appearance of
$6^n$ new knots at the $n$-th order of the Seiberg-Witten expansion.
These knots are trivial homology cycles which are Poincar\'e dual to
the high-order Seiberg-Witten potentials. Moreover the linking
number of a standard 1-cycle with the Poincar\'e dual of the gauge
field is shown to be written as an expansion of the linking number
of this 1-cycle with the Poincar\'e dual of the Seiberg-Witten gauge
fields. In the process we explicitly compute the noncommutative
'Jones-Witten' invariants up to first order in the noncommutative
parameter. Finally in order to exhibit a physical example, we apply
these ideas explicitly to the Aharonov-Bohm effect. It is explicitly displayed at first order in the
noncommutative parameter, we also show the relation to the noncommutative Landau
levels.

\begin{center}
\begin{minipage}[h]{14.0cm} {}
\end{minipage}
\end{center}

\vspace{2cm}


\bigskip

\noindent {\it Key words:}  Seiberg-Witten map; Wilson loop;
topological field theory; knot invariants.

\vspace{1cm}

\leftline{September 1, 2015}

\newpage

\section{Introduction}

Noncommutativity of spacetime has strongly attracted the attention
in the last two decades (see for instance
\cite{Douglas:2001ba,Szabo:2001kg}). In a remarkable paper N.
Seiberg and E. Witten \cite{SW}, they showed that open string theory
in the specific limit of small volume, non-vanishing $B$-field and
$\alpha ' \to 0$, the amplitudes with constant open string metric
$G$ and noncommutativity parameter $\Theta$, gives precisely a
(Connes-style \cite{Connes1}) noncommutative gauge theory. In this
context, this theory can be rewritten in terms of the commutative
one through a field redefinition known as the Seiberg-Witten map
\cite{SW}.

Under such a map gauge fields are written as an infinite series on
the noncommutative parameter $\Theta_{\mu \nu}$ ($[\widehat{x}_\mu,
\widehat{x}_\nu] = \Theta_{\mu \nu}$ where $\widehat{x}_\mu$ are the
noncommutative coordinates of the spacetime). To every order (in
$\Theta$) the gauge field can be determined in terms of the usual
(commutative) gauge field. The addition of higher-derivative terms
do not spoil gauge invariance since the noncommutative gauge group
action on the space of noncommutative connections, is such that the
quotient is isomorphic to the corresponding quotient in the
commutative case. This construction was realized explicitly for pure
gauge fields. Later this construction was extended to the non-abelian
case and coupled to matter fields \cite{JW1,JW2}. This proposal has
been studied widely in the literature and used for a noncommutative
gauge invariance extensions of the standard model and gravity (see
for instance, \cite{NCSM} and  \cite{GORS}). In the case of the
gravity such extension gave rise in a natural way to topological
invariants such as the Euler characteristic and the signature with
explicit computations in Ref. \cite{GORS2}.

It is well known that the Wilson lines and loops are very useful in
the description and computation of some non-perturbative aspects of
gauge theory just as confinement \cite{'tHooft:1977hy}. In the
context of string theory noncommutative Wilson loops have been
studied within the correspondence gauge/gravity duality in Refs.
\cite{Maldacena:1999mh,Ishibashi:1999hs,Dhar:2000nj,Huang:2007ux,Chakraborty:2012pc}.
In  Ref. \cite{Dhar:2001ff}, in noncommutative gauge theories,
Wilson lines were studied through the Schwinger-Dyson equations of
correlation functions of Wilson lines. In field theory on a
noncommutative two-dimensional torus the correlators of Wilson line
operators were determined \cite{Paniak:2003gn}. Also in the
two-dimensional plane within a perturbative (in $\Theta$ and in ${1
\over \Theta}$-expansion), the non-planar correlation functions of
Wilson loops were obtained and the mixing UV/IR was consistently
regularized \cite{Ambjorn:2004ck,Ambjorn:2007im}. Wilson loops also
were studied in the twisted covariant noncommutative field theory.
In particular their correlation functions, Morita duality and the
area preserving diffeomorphisms action, were examined in this
context in Refs. \cite{Cirafici:2005af,Riccardi:2007ku}. In Ref.
\cite{Bietenholz:2007ia}, it was observed that in a gauge theory on
the noncommutative plane, the area-preserving diffeomorphisms
symmetry is non-perturbatively broken.

Noncommutative gauge theories have very striking topological and
geometrical features. For instance, in string theory, in the absence
of $B$-field, the instanton equation $F^+=0$ is exact in $\alpha '$
as the instanton shrinks, and this small instanton becomes a
singularity of its moduli space. For non-vanishing $B$-field and
non-small volume, the singularity is resolved and absent from the
moduli space. This space defines precisely the moduli space of
noncommutative instantons \cite{Nekrasov:1998ss}. It is worth
mentioning that recently in Ref. \cite{AC} there were discussed some
geometrical implications of the Seiberg-Witten map in Chern-Simons
and gravity. Some useful comments of noncommutative Chern-Simons
theory can be found in \cite{SJ}. Some connections with the
Seiberg-Witten equations are described in Refs. \cite{GHI,HL,MW}.
The consideration of some topological aspects of the noncommutative
Wilson lines in the Seiberg-Witten limit is discussed in Ref.
\cite{Mukhi:2002gk}.  Another aspects of topological nature in
noncommutative gauge theory were discussed in Ref. \cite{harvey}.

The Wilson loops possesses by themselves some interesting
geometrical and topological properties e.g., in the abelian case,
they can be regarded in terms of the linking number \cite{RN}. On
the other hand Wilson loops are also useful for the description of
knots and link invariants. In Ref. \cite{WK} the Jones polynomials were reproduced and generalized by computing correlation functions
of products of Wilson loops using Chern-Simons action in the path
integral formalism. Essentially the computations throw topological
invariants since the action, observables and measure do not depends
on the metric for Chern-Simons theories and their higher-dimensional
generalization known as BF theories \cite{HS}. Linking numbers in BF
theories were examined in Refs. \cite{HS,GCSS,GPT}.

In the present work we explore some geometrical and topological
aspects based on the previous ideas but immersed in a noncommutative
space using the Wilson lines constructed by means of the gauge field provided by the Seiberg-Witten map. As our main result, we will see that in this context is
possible to establish a correspondence between the terms of a power series
(in the non-commutative parameter $\Theta$) series within the phase
of a non-commutative Wilson loop. Each term of the series in
$\Theta$ has associated various linking numbers, at the $n$-th term
of the expansion, there will arise $6^n$ extra linking numbers. All
these extra terms correspond to new homology cycles generated by the
non-vanishing parameter $\Theta$.

The linking number is ordinarily a topological invariant, now the
non-commutative linking numbers considered here, will represent a
topological invariant of the corresponding more general
non-commutative topology. Thus, the arising extra terms and their
involved mathematical structures, deserve a detailed mathematical and physical
interpretation and further analysis. In the present paper we will
restrict ourselves to compute the first order non-commutative
corrections to the linking numbers. This should be considered a
first attempt of a description of the subject.

To explore how the modifications to topology of knots immersed on the
non-commutative space we consider the abelian Jones-Witten
polynomials in the path integral formalism which are given in terms
of Wilson loops. We will show explicitly that the polynomials are
changed due to noncommutativity, at least up to first order. However we
should emphasize that, even at the first order, there will be a
non-vanishing and non-trivial modification of the linking numbers
due the noncommutative generalization of the notion of topology.

In order to explore our proposal in a more
detailed way we consider the application of the
non-commutative Wilson loops to the Aharonov-Bohm effect. Some
literature on the non-commutative Aharonov-Bohm effect and its
relation to the Landau levels can be found in Refs.
\cite{chaichian1,gamboa,chaichian2,Falomir,li,dulat}.

It is worth mentioning that the Wilson loops have been used in the
description of some quantum theories of gravity. Some of these
results can be found in Refs. \cite{rovelli,ashtekar, edwitten,
witten-kodama}. The results found in the present paper would be
applied also for this kind of theories.

This paper is structured as follows: in section $2$ we overview the
Seiberg-Witten map and setting up the notation and conventions that
we will follow along the paper. Section $3$ is devoted to construct
the noncommutative Wilson loops based on the gauge field of the Seiberg-Witten map. In sections $4$ we introduce the linking numbers, first in order to explore the
geometrical properties of noncommutative Wilson loops we use basic
ideas of Poincar\'e duality and interpret higher order powers in the
Seiberg-Witten expansion in terms of the arising new linking
numbers. In section $5$ we compute the abelian Jones-like
polynomials using the path integral formalism through the
Chern-Simons functional up to first order in the noncommutative
parameter.  As a physical application, in section $6$ we explore the
abelian noncommutative Aharonov-Bohm effect by means of the gauge field of the Seiberg-Witten map. The noncommutative Aharonov-Bohm effect is a
very interesting physical example in which the noncommutativity
could be relevant. It has known effects already described in the
literature \cite{chaichian1,gamboa,chaichian2,Falomir,li,dulat}.
Moreover it has a relation with the non-commutative Landau levels.
We will find that our results are consistent with these mentioned
results. Section 7 is devoted to final remarks.

\vskip 1truecm
\section{The Seiberg-Witten map}

Our aim is not to provide an extensive review on the Seiberg-Witten
map \cite{SW}, but only recall the relevant structure which will be
needed in the following sections. Throughout this paper we will
follow the notation and conventions introduced in reference
\cite{JW2}.

We are interested in noncommutativity utilizing the Seiberg-Witten
map \cite{SW}. This proposal was extended in \cite{JW1,JW2} for any
gauge field coupled to matter. Below we present a brief description
of this construction.

The central idea is to deform the algebraic structure of continuous
spaces in particular the polynomials in $N$ variables generated by
powers of $x^I$ where $I = 0, \ldots, N$, which is a freely
generated algebra $\mathbb{C}[x^1, \ldots, x^N]$.  Now consider the
usual commutations relations between the coordinates

\begin{equation}
[x^\mu, x^\nu ] = 0.
\end{equation}
This algebraic structure will be deformed assuming

\begin{equation}
[\widehat{x}^\mu, \widehat{x}^\nu ] = i \Theta^{\mu \nu},
\end{equation}
where $\Theta^{\mu \nu} = -\Theta^{\nu \mu} \in \mathbb{R}$ i.e.
$\Theta^{\mu \nu}$  is the noncommutative parameter and
$\widehat{x}$ are the noncommutative coordinates (as we can see this
algebra is similar to the Heisenberg algebra in the phase space) for
a formal description see \cite{JW2}.

Explicitly this modifies the way we multiply polynomials and in
general functions over the noncommutative variables in terms of the
commutative variables through the Moyal $\star$-product defined by:

\begin{equation}
f \star g (x) = \mu(e^{\frac{i}{2} \Theta^{\rho \sigma}
\partial_\rho \otimes \partial_\sigma}f \otimes g),
\end{equation}
where $\mu$ is the product map defined by $\mu(f(x) \otimes g(x)) =
f(x) \cdot g(x).$

In this context the gauge transformations of a matter field
$\Psi(x)$ is defined as
\begin{equation}
\label{var} \delta_{\Lambda} \Psi (x) = i \Lambda \star \Psi (x),
\end{equation}
$\Lambda(x) $ is the noncommutative gauge parameter which is Lie
algebra-valued i.e. $\Lambda (x) = \Lambda^a(x) T^a$. Now let us
compute explicitly the following variation of the field $\Psi$
$$
(\delta_{\Lambda_1} \delta_{\Lambda_2} - \delta_{\Lambda_1}
\delta_{\Lambda_2}) \Psi = (\Lambda_1 \star \Lambda_2 - \Lambda_2
\star \Lambda_1) \star \Psi
$$
\begin{equation}
\label{doublevar}
= \frac{1}{2} \left(  [\Lambda_1^a
\overset{\star}{,} \Lambda_2^b] \{T^a, T^b \} + \{ \Lambda_1^a
\overset{\star}{,} \, \Lambda_2^b \} [T^a, T^b ] \right) \star \Psi.
\end{equation}
It is worth mentioning that the fields are not Lie algebra-valued
because not only we have commutators but also we have
anticommutators. So the algebra that close both operations
(commutators and anticommutators) is precisely the Universal
Enveloping Algebra.

The covariant derivative is defined by
\begin{equation}
D_{\mu}^\star \Psi(x) = \partial_{\mu} \Psi(x) - i \widehat{A}_{\mu}
\star \Psi(x),
\end{equation}
where $\widehat{A}_{\mu}$ is the noncommutative gauge field and
transforms as

\begin{equation}
\label{varLambda} \delta_\Lambda \widehat{A}_\mu = \partial_\mu
\Lambda + i[ \Lambda \overset{\star}{,}\, \widehat{A}_\mu].
\end{equation}
We can see that these terms have infinite many degrees of freedom,
but in Ref. \cite{SW} it was shown that all the higher-order terms
depend only on the zeroth order terms (the commutative term) i.e. of
the gauge parameter $\Lambda^{(0) \, a} T^a$ and the gauge field
$A_\mu^{(0) \, a} T^a$. Let us assume that the gauge parameter
$\Lambda_\alpha$ depends only on $\alpha$ and $A_\mu$ i.e. the gauge
parameter and the gauge field respectively. With these assumptions
bearing in mind, let us substitute it in the expression
(\ref{doublevar})

\begin{equation}
\Lambda_\alpha \star \Lambda_\beta - \Lambda_\beta \star
\Lambda_\alpha +i(\delta_\alpha \Lambda_\beta - \delta_\beta
\Lambda_\alpha) = i \Lambda_{-i[\alpha, \beta]}.
\end{equation}
This expression could be solved perturbatively assuming an expansion
in the parameter $\Theta$ as $\Lambda_\alpha = \Lambda_\alpha^{(0)}
+ \Lambda_\alpha^{(1)} + \cdots$, where $\Lambda_\alpha^{(0)} =
\alpha = \alpha^a T^a$.

For example up to first order in $\Theta$, we find the following
expression:

 \begin{eqnarray}
 \delta_\alpha \Lambda_\beta^{(1)} - \delta_\beta \Lambda_\alpha^{(1)} -
 i [\alpha, \Lambda_\beta^{(1)}] -i[\Lambda_\alpha^{(1)}, \beta]- \Lambda_{-i[\alpha,
 \beta]}^{(1)}
 = - \frac{1}{2} \Theta^{\mu \nu} \{ \partial_\mu \alpha, \partial_\nu \beta \},
\end{eqnarray}
whose solution is $\Lambda_{\alpha}^{(1)} = \alpha - \frac{1}{4}
\Theta^{\mu \nu} \{ A_\mu, \partial_\nu \alpha \}$. With this
expression we can compute in a similar way  the expression for
matter fields assuming the transformation at zero-th order
$\delta_\alpha \Psi^{(0)} = i \alpha \Psi^{(0)}$.

For the gauge field we expand again in orders of $\Theta$ as
$\widehat{A} = A^{(0)} + A^{(1)} + A^{(2)} + \cdots$ and
substituting it into Eq. (\ref{varLambda}), up to first order in
$\Theta$, we obtain

 \begin{equation}
 A_\mu^{(1)} = \frac{1}{4} \Theta^{\rho \sigma}(\{ F_{\rho \mu}, A_\sigma\}
 - \{A_\rho, \partial_\sigma A_\mu\}).
 \end{equation}
 Finally the field strength tensor is given by $F_{\mu \nu}^\star
 = i [D_\mu^\star \, \overset{\star}{,} \, d_\nu^\star]$, whose solution up to first order is

 \begin{eqnarray}
 \widehat{F}_{\mu \nu} = F_{\mu \nu} + \frac{1}{4} \Theta^{\rho \sigma}(2 \{ F_{\rho \mu}, F_{\sigma \nu}\}
 + \{ D_\rho F_{\mu \nu}, A_{\sigma}\} - \{ A_{\rho}, \partial_\sigma F_{\mu \nu}\}).
 \end{eqnarray}
Here we can identify $F_{\mu \nu}^{(0)} = F_{\mu \nu}$ and $F_{\mu
\nu}^{(1)} =  \frac{1}{4} \Theta^{\rho \sigma} (2 \{ F_{\rho \mu},
F_{\sigma \nu}\} + \{ D_\rho F_{\mu \nu}, A_{\sigma}\} - \{
A_{\rho},
\partial_\sigma F_{\mu \nu}\})$.

\vskip 2truecm
\section{Noncommutative Wilson Loops}

The usual Wilson loop is given by the following expression,

\begin{equation}
W(C) = \Tr P \exp \left( \frac{i}{\hbar} \int_C A \right),
\end{equation}
where $A=A_\mu^a t^a dx^\mu$, with $t^a$ being the Lie algebra
generators, then the Wilson loop is an element of the Lie group  and
an element of the holonomy. An extension to the noncommutative case
 was proposed in a straightforward manner using the Moyal product in\cite{GHI,MW}. In this work we make use of the gauge field given by the Seiberg-Witten map in order to construct the corresponding noncommutative  Wilson loop, this amounts to change the connection $A$ by $\widehat{A}$ which possesses an
expansion in powers of $\Theta$ given by $\widehat{A} = A^{(0)} +
A^{(1)} + A^{(2)} + \cdots$ and every $A^{(i)}$ is expanded in terms
of the usual $1$-form basis i.e $A^{(i)}= A^{(i)}_\mu dx^\mu$, the Wilson loop corresponding to the Seiberg-Witten map is given by

\begin{equation}
\label{ncwilson} \widehat{W}(C) = \Tr P \exp_{\star} \left(
\frac{i}{\hbar} \int_C \widehat{A} \right),
\end{equation}
For Chern-Simons (our case of interest) it was shown in Ref.
\cite{AC} that the same formulae apply for the Seiberg-witten map.
Further developments for higher-order computations of the
Seiberg-Witten map can be found in  \cite{UY}.

The $\star$-exponential is defined as
\begin{multline}
\Tr \exp_\star \left (\frac{i}{\hbar} \int_C \widehat{A} \right) = 1
+ \frac{i}{\hbar}\int_C \Tr \widehat{A}(x)  \\ + \frac{i^2}{\hbar^2
2!}\int_C \int_C \Tr \left( \widehat{A}(x) \star \widehat{A}(y)
\right) + \frac{i^3}{\hbar^3 3!} \int_C \int_C \int_C \Tr \left(
\widehat{A}(x) \star \widehat{A}(y) \star \widehat{A}(z) \right) +
\cdots,
\end{multline}
where we are assuming the integration order $x < y < z < \cdots$,
with $x, y, z, \ldots \, \in \IC$.

\subsection{Abelian Case}
For the abelian case the Wilson loop can be written as follows:

\begin{equation}
\label{NC-Wilson} \widehat{W}(C) = \exp_\star \left( {i \over \hbar}
\int_C \widehat{A} \right) =\exp_\star \left( {i \over \hbar} \int_C
A^{(0)} \right) \exp_\star \left( {i \over \hbar} \int_C A^{(1)}
\right)\exp_\star \left( {i \over \hbar} \int_C A^{(2)} \right)
\cdots \ \ .
\end{equation}
Then we can compute explicitly the Wilson loop up to second $\Theta$
order. We can see that it is necessary to expand the first three
exponentials

\begin{equation}
\widehat{W}(C) \approx \exp_\star \left( {i \over \hbar} \int_C
A^{(0)} \right) \exp_\star \left( {i \over \hbar} \int_C A^{(1)}
\right) \exp_\star \left( {i \over \hbar} \int_C A^{(2)} \right).
\end{equation}
It is easy to check the exponential $\exp_\star \left( {i \over
\hbar} \int_C A^{(j)} \right)$ up order $2j +1$ is given by

\begin{multline}
\exp_\star \left( {i \over \hbar} \int_C A^{(j)} \right) = \\ 1+{i
\over \hbar} \int_C A^{(j)} + \frac{i^2}{2!\hbar^2} \int_{C \times
C} A^{(j)} \star A^{(j)} + \frac{i^3}{3!\hbar^3} \int_{C \times C
\times C} A^{(j)} \star A^{(j)} \star A^{(j)} + \cdots \\ \approx
\exp \left( {i \over \hbar} \int_C A^{(j)} \right) \left[ 1 +
\frac{i^2}{2! \hbar^2} \frac{i}{2} \Theta^{\mu \nu}
\partial_\mu \int_C A^{(j)} \cdot \partial_\nu \int_C A^{(j)}
\right].
\end{multline}
The second term vanishes in virtue of the abelianity and the skew
symmetry of $\Theta^{\mu \nu}$, thus the second order Wilson loop is
given by

\begin{eqnarray}
\label{2orderWL}
\widehat{W}(C) \approx \exp \left( \frac{i}{\hbar} \int_C A^{(0)} \right) \left[ 1
+ \frac{i}{\hbar} \int_C A^{(1)} + \frac{i}{\hbar} \int_C A^{(2)} + \frac{1}{2!} \left( \frac{i}{\hbar} \right)^2 \int_C A^{(1)} \int_C A^{(1)} \right] \nonumber \\
= \exp \left( \frac{i}{\hbar} \alpha_0 \right) \left[ 1
+ \frac{i}{\hbar} \Theta \alpha_1 + \frac{i}{\hbar} \Theta^2 \alpha_2 + \frac{1}{2!} \left( \frac{i}{\hbar} \right)^2 \Theta^2 \alpha_1 \alpha_1 \right].
\end{eqnarray}

As an example let us compute $\widehat{W}(C)$ assuming a pure
constant magnetic field a long the $z$-axis $\vec{B} = B_0
\widehat{k} $ and $C=S^1$ in the $x-y$ plane, hence $A_1^0 =
-\frac{B_0}{2}y$ and $A_2^0 = \frac{B_0}{2}x$, thus

\begin{eqnarray}
\label{magnetic}
\alpha_0 = \int_C A^{(0)}_\mu dx^\mu = \frac{B_0}{2}(2 \pi) = \phi, \\
\Theta \alpha_1 = \int_C A^{(1)}_\mu dx^{\mu}= \frac{3}{4}B_0^2 (2
\pi) \Theta^{12} = \frac{3}{2 \pi} \phi^2 \Theta^{12} ,\\ \Theta^2
\alpha_2 = \int_C A^{(2)}_\mu dx^\mu = \frac{1}{2}B_0^3 (2 \pi)
(\Theta^{12})^2 = \frac{4}{(2 \pi)^2} \phi^3 (\Theta^{12})^2,
\end{eqnarray}
where $\phi$ is the flux due to the commutative field through the
closed curve $C$. Now up to second order the noncommutative Wilson
line is given by

$$
\widehat{W}(C) \approx \exp(\frac{i}{\hbar} \alpha_0) \left( 1 + \Theta \frac{i}{\hbar} \alpha_1
+ \Theta^2 \left( \frac{i}{\hbar} \alpha_2 + \frac{1}{2} \left( \frac{i}{\hbar} \right)^2 \alpha_1^2 \right) \right)
$$
\begin{equation}
= \exp \left( \frac{i}{\hbar} \frac{B_0}{2}(2 \pi) \right) \left( 1 + \frac{i}{\hbar}\Theta^{12}
\left[ \frac{3}{4}B_0(2 \pi) \right] + \frac{1}{2}\frac{i}{\hbar}( \Theta^{12})^2
\left[ (2 \pi) B_0^3 +\frac{i}{\hbar}(\frac{3}{4}B_0^2)^2(2 \pi)^2 \right] \right).
\end{equation}
Therefore as we might expect, to observe noncommutativity effects
assuming $\Theta \ll 1$ we need the intensity of the magnetic field
$B_0$ to be large enough.


\section{Linking Numbers and the Wilson Loops}

It is well known that on a three manifold $M= S^3$ or $\mathbb{R}^3$
the magnetic field induced by a loop wire carrying a current, is
proportional to the linking number between the loop and the magnetic
lines, due the Biot-Savart law (which is essentially the Gauss
linking number). This is written usually as

\begin{equation}
\label{cflux} \Phi = \int_\Sigma F^{(0)} = \int_\Sigma dA^{(0)} =
\int_C A^{(0)},
\end{equation}
where we assuming that $C$ is a trivial homology 1-cycle (i.e. a
trivial element in the first homology group). Thus there exists a
$2$-chain $\Sigma$ (it could be regarded as a two-dimensional
submanifold of $\mathbb{R}^3$) such that $C= \partial \Sigma$ is
immersed on the three-dimensional Euclidean space. The connection
$A^{(0)} = A_\mu^{(0)} dx^\mu$ is the commutative (usual) magnetic
potential $1$-form. As we can see the last integral is the phase
that appears on the commutative Wilson loop.

Since $C$ is a trivial cycle by Poincar\'e duality there exists $d
\eta$ being $\eta$ a $1$-form. The mathematical interpretation
assuming that $C$ is a trivial $1$-cycle and the flux is non-zero
implies that $A^{(0)}$ is not closed form, i.e.

\begin{equation}
\Phi = \int_M A^{(0)} \wedge d \eta = \int_M \eta \wedge d A^{(0)}.
\end{equation}

In the previous integral we integrate out by parts in this way we
can regard $d A^{(0)}$ as the Poincar\'e dual of some trivial cycle
$C^{(0)}$ which is boundary of the $2$-chain $\Sigma^{(0)}$. Hence
this integral represents the linking number among the trivial cycles
$C$ and $C^{(0)}$ being them the respective boundaries of the
$2$-cycles $\Sigma$ and $\Sigma^{(0)}$.

Now we will consider the noncommutative case, then the phase in the
noncommutative Wilson loop assuming the expansion on $\Theta$ is
given by

\begin{equation}
\int_C \widehat{A} =  \int_C A^{(0)} + \int_C A^{(1)} + \int_C
A^{(2)} + \cdots   \ \ .
\end{equation}
Interpreting this in terms of linking numbers, the first term is the
link number between $C$ and $C^{(0)}$, the second term is
interpreted in the following way. Let us consider from the
Seiberg-Witten map the first order potential $A_\mu^{(1)} =
-\Theta^{\kappa \lambda} \frac{1}{2} A_\kappa^{(0)}
(\partial_\lambda A_\mu^{(0)} + F_{\lambda \mu}^{(0)})$ that can be
arranged as

\begin{equation}
\label{1potential} A^{(1)} = \Theta^{\kappa \lambda} A_{\kappa
\lambda}^{(1)} = \Theta^{\kappa \lambda} A_{\kappa \lambda
\mu}^{(1)} dx^\mu.
\end{equation}
In this way $A^{(1)}$ can be regarded as the sum of six $1$-forms
$A^{(1)}_{\kappa \lambda} = A^{(1)}_{\kappa \lambda \mu} dx^\mu$
(since we are working in $\mathbb{R}^3$)

\begin{equation}
\label{1order} \int_C A^{(1)} = \int_M A^{(1)} \wedge d \eta =
\int_M \eta \wedge d A^{(1)} = \int_{C^{(1)}} \eta,
\end{equation}
where $C^{(1)}$ is a trivial $1$-cycle. But explicitly the first
integral $\Theta^{\kappa \lambda} \int_C A^{(1)}_{\kappa \lambda}$
must be equal to $\int_{C^{(1)}} \eta$ therefore it also will be
expanded in $\Theta$ then Eq. (\ref{1order}) reads

\begin{equation}
\Theta^{\kappa \lambda} \int_C A^{(1)}_{\kappa \lambda} =
\Theta^{\kappa \lambda}\int_{C^{(1)}_{\kappa \lambda}} \eta.
\end{equation}
This equation implies that $C^{(1)}_{\kappa \lambda} = \partial
\Sigma^{(1)}_{\kappa \lambda}$ is a trivial cycle, thus it can be
regarded as the linking number between $C$ and each $C_{\kappa
\lambda}^{(1)}$, e.g. in the three dimensional Euclidean space

\begin{equation}
\Theta^{\kappa \lambda}\int_{C^{(1)}_{\kappa \lambda}} \eta =
\Theta^{12} \bigg(\int_{C^{(1)}_{12} } \eta -\int_{C^{(1)}_{21}}
\eta \bigg) + \Theta^{13} \bigg(\int_{C^{(1)}_{13} } \eta
-\int_{C^{(1)}_{31}} \eta \bigg) + \Theta^{23}
\bigg(\int_{C^{(1)}_{23} } \eta -\int_{C^{(1)}_{32}} \eta \bigg),
\end{equation}
or

\begin{equation}
\Theta^{\kappa \lambda} \int_C A^{(1)}_{\kappa \lambda} =
\Theta^{12} \int_{C} (A_{12}^{(1)} - A_{21}^{(1)}) + \Theta^{13}
\int_{C} (A_{13}^{(1)} - A_{31}^{(1)}) + \Theta^{23} \int_{C}
(A_{23}^{(1)} - A_{32}^{(1)}),
\end{equation}
in general since $A_{\kappa \lambda}^{(1)} \neq A_{\lambda
\kappa}^{(1)}$, then the linking number between $C$ and
$C^{(1)}_{\kappa \lambda}$ is different from the one of $C$ and
$C^{(1)}_{\lambda \kappa}$. In figure (1) it is explicitly displayed
the intersection between the  $C^{(1)}_{\lambda \kappa}$ 1-cycles,
represented in different colors by the simplest case when they are
circles, and the original 1-cycle $C$.

\begin{figure}
\begin{center}
\vskip -1.5truecm
\includegraphics[scale=0.5]{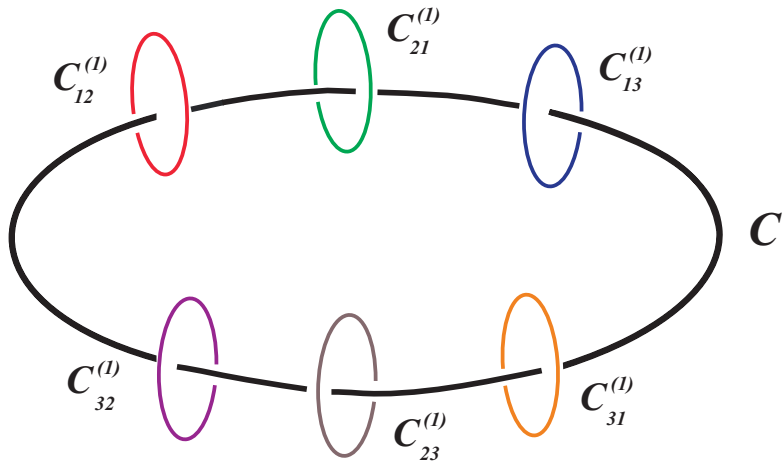}
\vskip -1.5truecm \caption[Short caption for figure
1]{\label{fig:thetaknots} {\small The figure accounts the computation
at the first order in $\Theta$ of the linking number. For the case
of knots in $\mathbb{R}^3$ it is observed at this order that the
noncommutativity induces the existence of 6 different knots,
represented in the simplest case by circles of different colors
$C^{(1)}_{\kappa \lambda}$, that intersect the original trivial
homology $1$-cycle $C$. At the $n$-th order there will be $6^n$
knots intersecting $C$. In the general case the knots
$C^{(1)}_{\kappa \lambda}$ would be truly knots, as for instance the
trefoil or even more complicated knots.}}
\end{center}
\end{figure}

Let us introduce the following notation: $\rm{Lk}(C_1, C_2)$ denotes
the linking number between the homologically trivial $1$-cycles
$C_1$ and $C_2$.

Now consider the second order term and reordering the expression
$A^{(2)}$ in the Seiberg-Witten map, it could be written in a
similar way as in Eq. (\ref{1potential})

\begin{equation}
A^{(2)} = \Theta^{\kappa_1 \lambda_1} \Theta^{\kappa_2 \lambda_2}
 A^{(2)}_{\kappa_1 \lambda_1 \kappa_2 \lambda_2 \, \mu} dx^\mu.
\end{equation}
Let $A^{(2)}_{\kappa_1 \lambda_1 \kappa_2 \lambda_2} =
A^{(2)}_{\kappa_1 \lambda_1 \kappa_2 \lambda_2 \, \mu} dx^\mu$, in
analogy to the first order term

\begin{equation}
\Theta^{\kappa_1 \lambda_1} \Theta^{\kappa_2 \lambda_2} \int_C
A^{(2)}_{\kappa_1 \lambda_1 \kappa_2 \lambda_2} = \Theta^{\kappa_1
\lambda_1} \Theta^{\kappa_2 \lambda_2} \int_{C^{(2)}_{\kappa_1
\lambda_1 \kappa_2 \lambda_2}} \eta,
\end{equation}
again here $d \eta$ denotes is de Poincar\'e dual of $\gamma$. As we
can see in general the total flux induced by the second order term
is the sum of the $36$ linking numbers $\rm{Lk}(C, C^{(2)}_{\kappa_1
\lambda_1 \kappa_2 \lambda_2})$.

Finally in general for the $m$-th order we rearrange the expression
$A^{(m)}$ from the Seiberg-Witten map we rewrite it as:

\begin{equation}
A^{(m)} = \Theta^{\kappa_1 \lambda_1} \cdots \Theta^{\kappa_m
\lambda_m} A^{(m)}_{\kappa_1 \lambda_1 \cdots \kappa_m \lambda_m \,
\mu} dx^\mu.
\end{equation}
We define $ A^{(m)}_{\kappa_1 \lambda_1 \cdots \kappa_m \lambda_m} =
A^{(m)}_{\kappa_1 \lambda_1 \cdots \kappa_m \lambda_m \mu} dx^\mu$.
Thus we have

\begin{multline}
\Theta^{\kappa_1 \lambda_1} \cdots \Theta^{\kappa_m \lambda_m}
\int_C A^{(m)}_{\kappa_1 \lambda_1 \cdots \kappa_m \lambda_m} =
\Theta^{\kappa_1 \lambda_1} \cdots \Theta^{\kappa_m \lambda_m}
\int_{C^{(m)}_{\kappa_1 \lambda_1 \cdots \kappa_m \lambda_m}} \eta   \\
= \Theta^{\kappa_1 \lambda_1} \cdots \Theta^{\kappa_m \lambda_m}
\rm{Lk}(C, C^{(m)}_{\kappa_1 \lambda_1\cdots \kappa_m \lambda_m}),
\label{newknots}
\end{multline}
where $C^{(m)}_{\kappa_1 \lambda_1\cdots \kappa_m \lambda_m}$ are
homologically trivial $1$-cycles i.e. exist a $2$-chain
$\Sigma^{(m)}_{\kappa_1 \lambda_1\cdots \kappa_m \lambda_m}$, such
that $C^{(m)}_{\kappa_1 \lambda_1\cdots \kappa_m \lambda_m} =
\partial \Sigma^{(m)}_{\kappa_1 \lambda_1\cdots \kappa_m \lambda_m}$.
Therefore in general the $m$-flux through $C$ could be regarded as
the sum of $6^m$ linking numbers between $C$ and $C^{(m)}_{\kappa_1
\lambda_1\cdots \kappa_m \lambda_m}$ in $\mathbb{R}^3$.

\section{The Jones-Witten-Like Invariants}

The Jones polynomials in the Witten's path integral formulation are
given  by the correlation functions of Wilson loops
\begin{equation}
J_C = \langle {W}(C) \rangle = \frac{1}{N}\int \mathcal{D} A \exp
\left[ \frac{i}{\hbar}k \int_{M} \Tr \bigg( A \wedge dA + A \wedge A
\wedge A \bigg) \right] {\Tr}_R \,W(C),
\end{equation}
where $W(C) = P \exp[ \frac{i}{\hbar}\int_C A]$ is the Wilson line,
$R$ is a representation of the gauge group, $A$ is the gauge field,
$C$ is a knot (homologically trivial $1$-cycle), $M$ the Euclidean
$3$-dimensional space or the unitary $3$-sphere and $N$ is the
normalization factor.

We will consider the simplest case when the gauge group is $U(1)$
but consider the noncommutative Wilson loop (abelian case), and
compute the first non-trivial correction which will arise at first
order in $\Theta$.

First of all we substitute in the path integral $A$ by
$\widehat{A}$, i.e the noncommutative Jones-Witten like invariants will depend explicitly of the noncommutative parameter (for
the abelian case) read,
\begin{equation}
J_C(\Theta) =
\langle \widehat{W}(C) \rangle = \frac{1}{N}\int \mathcal{D}
\widehat{A} \exp \left[\frac{i}{\hbar} k \int_{M} \Tr \bigg(
\widehat{A} \overset{\star}{\wedge} d \widehat{A} + \widehat{A}
\overset{\star}{\wedge} \widehat{A} \overset{\star}{\wedge}
\widehat{A} \bigg) \right] \rm{Tr}_R \,\widehat{W}(C).
\end{equation}
Since we will consider the $\Theta$-expansion just up to first
order, the term $\widehat{A} \overset{\star}{\wedge} \widehat{A}
\overset{\star}{\wedge} \widehat{A}$ does not contribute (it
contributes up to third order), then we get the following expression
for the action

\begin{multline}
\exp \left[ \frac{i}{\hbar}k\int_M (A^{(0)} \wedge dA^{(0)} + 2 A^{(0)} \wedge d A^{(1)}) \right] \\
 \approx \exp \left[ \frac{i}{\hbar}k\int_M A^{(0)} \wedge d A^{(0)} \right]
 \left[ 1+ \frac{i}{\hbar} 2 k \Theta^{\kappa \lambda} \int_M A^{(0)} \wedge dA^{(1)}_{\kappa
\lambda} \right],
\end{multline}
and the noncommutative Wilson line (\ref{ncwilson}) up to first
order in $\Theta$ is

\begin{equation}
\widehat{W}(C) = \exp_{\star} \left(\frac{1}{\hbar} \int_C \widehat{A} \right)\approx  \exp \left( \frac{i}{\hbar} \int_C A^{(0)}
\right) \left[ 1 + \frac{i}{\hbar} \Theta^{\kappa \lambda} \int_C
A^{(1)}_{\kappa \lambda} \right].
\end{equation}
Then the functional integral up to first order (the first $\Theta$
order Jones polynomials) is given by

\begin{multline}
\int \mathcal{D} A^{(0)}  \exp \left( \frac{i}{\hbar}k \int_{M}
A^{(0)} \wedge dA^{(0)}
\right)  \\
\times \exp \left( \frac{i}{\hbar} \int_C A^{(0)} \right)  \left[ 1
+ \frac{i}{\hbar} \Theta^{\kappa \lambda} \int_C A^{(1)}_{\kappa
\lambda} + \frac{i}{\hbar} 2 k \Theta^{\kappa \lambda} \int_M
A^{(0)} \wedge dA^{(1)}_{\kappa \lambda} \right].
\end{multline}
The integral is performed just on the commutative term $A^{(0)}$
since every order can be written in term $A^{(0)}$ in virtue of
Seiberg-Witten map. Rewriting the previous expression we get

\begin{multline}
\label{Jones.1order} \int \mathcal{D} A^{(0)}  \exp \left(
\frac{i}{\hbar}k \int_{M} (A^{(0)} \wedge dA^{(0)} +
\frac{1}{k}  A^{(0)} \wedge d \eta ) \right)  \\
\times \left[ 1 +\frac{i}{\hbar} \Theta^{\kappa \lambda} \int_M
A^{(1)}_{\kappa \lambda} \wedge d \eta + \frac{i}{\hbar} 2 k
\Theta^{\kappa \lambda} \int_M A^{(0)} \wedge dA^{(1)}_{\kappa
\lambda} \right],
\end{multline}
where $d \eta$ is the Poincar\'e dual of the knot $C$. Next we need
to compute these three integrals; the first one is the usual abelian
Jones polynomial like for the knot $C$, whose evaluation is
proportional to

\begin{equation}
\exp \bigg(-{i \over 4 \hbar k} \int_M \eta \wedge d \eta \bigg),
\label{arbol}
\end{equation}
where $\int_M \eta \wedge d \eta$ is the Hopf invariant or
self-linking number.

Now we consider the second integral which is rewritten as
\begin{eqnarray}
\label{1Path}
\frac{i}{\hbar} \Theta^{\kappa \lambda} \int \mathcal{D} A^{(0)}
\exp \left[\frac{i}{\hbar}k \int_{M} (A^{(0)} \wedge dA^{(0)} +
\frac{1}{k} A^{(0)} \wedge d \eta) \right] \left[ \int_M
A^{(1)}_{\kappa \lambda} \wedge d \eta \right].
\end{eqnarray}
First of all, notice that  any $1$-form $A^{(0)}$, using the Hodge
decomposition theorem, can be uniquely decomposed as $A^{harm} + d B
+ \delta C$, where $A^{harm}$ is a harmonic $1$-form (i.e. it is a
solution to the laplacian $\Delta_p=d \delta + \delta d$), $B$ is a
$0$-from, and $C$ is a $2$-form. Here $d$ the usual exterior
derivative (which maps a $p$-form into a $(p+1)$-form) and $\delta =
(-1)^{n(p+1)+1} \ast d \ast$ their adjoint operator (which maps
$p$-form into $(p-1)$-form). The $\ast$ stands for the Hodge star
operator which maps a $p$-form into an $(n-p)$-form. Also it is
necessary to consider that $\ast^2 = (-1)^{p(n-p)}$.\\

In order to integrate out the path integral measure is decomposed
into $\mathcal{D} A^{(0)} = \mathcal{D} A^{harm} \mathcal{D} A^t
\mathcal{D} A^{l}$ in virtue of the Hodge decomposition theorem,
where $A^t = d B$ and $A^{l} = \delta C$ are the longitudinal and
transversal parts of $A^{(0)}$ respectively. It is well known that
just the transversal parts contributes to the integration.

Moreover it is necessary to introduce some extra conventions. Let
$\Lambda_p$ the space of eigenforms of $\Delta_p$ with non-vanishing
eigenvalue $\lambda^2$. This space can be decomposed into
$\Lambda_p^l$ and $\Lambda_p^t$ i.e. into the longitudinal and
transversal parts. We also need to consider the following maps:
$d:\Lambda_{p}^t \to \Lambda_{p+1}^l$, $\delta:\Lambda_{p+1}^l \to
\Lambda_{p}^t$ and an isomorphism given by $\ast: \Lambda_p \to
\Lambda_{n-p}$ among forms with the same eigenvalue $\lambda^2$.
Given these maps finally we construct the following isomorphism:

\begin{equation}
\label{iso}
\lambda^{-1} \ast d: \Lambda_p^t \to \Lambda_{n-p-1}^t.\\
\end{equation}
Since we will consider $M =S^3$ or $\mathbb{R}^3$, and $1$-forms
(this is $n=3$ and $p=1$), the last isomorphism is reduced to
$\lambda^{-1} \ast d: \Lambda_1^t \to \Lambda_{1}^t$.\\

Now consider a basis of transverse $1$-eigenforms $\{A_i^t\}$ such
that it satisfies the normalization condition $\langle A^t_i | A^t_j
\rangle = \int_M A^t_i \wedge \ast A^t_j = \delta_{ij}$ (inner
product among $1$-forms). Hence we expand the transversal part of
$A^{(0)}$ and $\eta$ in term of this basis hence $A^t = \sum_i
a^{(0)}_i A_i^t$ and $\eta^t = \sum_i \eta_i A_i^t$.  With the aid
of the inner product, $\ast^2 = +1$ for $2$-forms in a Riemannian
manifold and the isomorphism (\ref{iso}) the term in the exponential
of the left-hand-side of (\ref{1Path}) can be rewritten as follows

\begin{equation}
 k \langle A^{(0)} | \ast dA^{(0)} \rangle + \langle A^{(0)} | \ast d \eta
\rangle = k \sum_i \lambda_i \left((a^{(0)}_i)^2 + \frac{1}{k}
a^{(0)}_i \eta_i \right),
\end{equation}
where $\ast d A^{(0)} = \ast d \sum_i a^{(0)}_i A^t_i = \sum_i
a^{(0)}_i  \ast d A^t_i = \sum_i a^{(0)}_i \lambda_i A^t_i$ and
similarly for $\ast d\eta$.

In a similar spirit we compute the first order $\Theta$ expansion of
the Wilson line
\begin{equation}
\label{1expantion} \Theta^{\kappa \lambda} \int_M A^{(1)}_{\kappa
\lambda} \wedge d \eta = \int_M A^{(1)} \wedge d \eta = \sum_i
\lambda_i a^{(1)}_i \eta_i,
\end{equation}
being $A^{(1)} = \sum a^{(1)}_\ell A^t_\ell$. Explicitly
$a^{(1)}_\ell$ can be written in terms of $a^{(0)}$ since
$A^{(1)}_\mu = -\frac{1}{2} \Theta^{\kappa \lambda} A^{(0)}_\kappa
(\partial_\lambda A^{(0)}_\mu + F^{(0)}_{\lambda \mu})$. Using the
eigenbasis expansion $A^{(0)} = \sum_i a^{(0)}_i A_i^t$ and the
following relation

\begin{equation}
d A_i^t =\ast^2 d A_i^t= \ast \lambda_i A_i^t = \frac{1}{2}
\lambda_i  A_{i \mu}^t \varepsilon^\mu_{ \ \rho \sigma} dx^{\rho}
\wedge dx^{\sigma},
\end{equation}
where we use the fact $\ast^2 = 1$. In virtue of the last expression
we get the following useful expression for the partial derivatives
of the gauge field components $\partial_\rho A_{i \sigma} = \frac{1}{2} \lambda_i  A_{i \mu}^t \varepsilon^\mu_{ \ \rho
\sigma}$, hence we can rewrite $A^{(1)}$ as:

\begin{equation}
A^{(1)} = \frac{3}{4} \Theta^{\kappa \lambda} \sum_{i,j} \lambda_j
a_i^{(0)} a_j^{(0)}  A_{i \kappa}^t A_{j \mu}^t \varepsilon^\mu_{ \
\rho \lambda} d x^\rho.
\end{equation}
Finally projecting out $A^{(1)}$ into the transverse basis we get
the following explicitly expression
\begin{equation}
\label{1coef} a_\ell^{(1)} = \int_M A^{(1)} \wedge \ast A_{\ell} =
\frac{3}{8} \Theta^{\kappa \lambda} \sum_{i,j} \lambda_j a^{(0)}_i
a^{(0)}_j [C_{ij\ell}]_{\kappa \lambda},
\end{equation}
where the $C$'s are defined by the following expression

\begin{equation}
[C_{ij\ell}]_{\kappa \lambda} = \int_M \varepsilon^{\mu}_{ \ \gamma
\lambda} \varepsilon^{\zeta}_{ \  \chi \rho} A^t_{i \kappa} A^t_{j
\mu} A^t_{\ell \zeta} \cdot dx^\gamma \wedge d x^{\chi} \wedge d
x^{\rho}.
\end{equation}
With these algebraic manipulations we can write Eq.
(\ref{1expantion}) as

\begin{equation}
\int_M A^{(1)} \wedge d \eta = \frac{3}{8} \Theta^{\kappa \lambda}
\sum_{i,j,\ell}  \eta_\ell  \lambda_j \lambda_\ell  a^{(0)}_i
a^{(0)}_j [C_{ij \ell}]_{\kappa \lambda}.
\end{equation}
Following a similar procedure we find

\begin{equation}
\int_M A^{(0)} \wedge d A^{(1)} = \frac{3}{8} \Theta^{\kappa
\lambda} \sum_{i,j,\ell} \lambda_j \lambda_\ell  a^{(0)}_i a^{(0)}_j
a^{(0)}_\ell [C_{ij \ell}]_{\kappa \lambda}.
\end{equation}

Therefore Eq. (\ref{1Path}) can be re-expressed in terms of the
previous expansion then

\begin{multline}
\label{1pathexpanded} \frac{1}{N}\int \mathcal{D} A^{(0)} \exp
\left[\frac{i}{\hbar}k \int_{M} (A^{(0)} \wedge dA^{(0)} +
\frac{1}{k} A^{(0)} \wedge d \eta) \right] \cdot
\left[ \frac{i}{\hbar}  \int_M A^{(1)} \wedge d \eta \right] \\
= \frac{1}{N} \int \prod_m d a^{(0)}_m \exp \left[ \frac{i}{\hbar}
k\left( \sum_m \lambda_m [(a^{(0)}_m)^2 + \frac{1}{k} a^{(0)}_m
\eta_m] \right)\right]
\\
\times \left[ \frac{3}{8} \frac{i}{\hbar} \Theta^{\kappa \lambda}
\sum_{i,j,\ell} \eta_\ell  \lambda_j \lambda_\ell a_i^{(0)}
a^{(0)}_j
 [C_{ij \ell}]_{\kappa \lambda} \right],
\end{multline}
where $N = \int \mathcal{D} A^{(0)} \exp \left[ik \int_{M} A^{(0)}
\wedge dA^{(0)}  \right]$ is a normalization factor which is given
by

\begin{eqnarray}
N =\lim_{n \to \infty} \left( \frac{i \hbar \pi}{
k}\right)^{\frac{n}{2}} \frac{1}{\sqrt{\det \Delta}},
\end{eqnarray}
where $\det \Delta = \prod_{m=1}^n \lambda_m.$\\

In order to integrate out the expression (\ref{1pathexpanded}), is
convenient to rewrite it as follows

\begin{equation}
\frac{i}{\hbar} \Theta^{\kappa \lambda} \frac{3}{8} \sum_{\ell, m,
i} \lambda_\ell \eta_i \lambda_i [C_{\ell mi}]_{\kappa \lambda} \int
\prod_n da_n^{(0)} \exp \left( -\frac{1}{2} \sum_{i,j} a_i^{(0)}
A_{ij} a_j^{(0)} + \sum_i a_i^{(0)} J_i \right) a^{(0)}_\ell
a^{(0)}_m,
\end{equation}
where $A_{ij} = - \frac{2i}{\hbar} k \lambda_j \delta_{ij}$ and $J_i
= \frac{i}{\hbar} \lambda_i \eta_i$, integrating out we obtain

%

\begin{multline}
\frac{1}{N}\int \mathcal{D} A^{(0)} \exp \left[\frac{i}{\hbar} k
\int_{M} (A^{(0)} \wedge dA^{(0)} + \frac{1}{k} A^{(0)} \wedge d
\eta) \right] \left[ i  \int_M A^{(1)} \wedge d \eta \right] \\ =
\frac{3}{8} \Theta^{\kappa \lambda} \sum_{i,j,\ell} \lambda_j
\lambda_\ell [C_{i j \ell}]_{\kappa \lambda} \exp \left( -
\frac{i}{4k \hbar} \int_M \eta \wedge d\eta \right) \left(
\frac{i}{4 \hbar k^2} \eta_i  \eta_j \eta_{\ell}- \frac{1}{2 k}
\frac{\delta_{i j} \eta_\ell}{\lambda_j}  \right).
\end{multline}

Finally the third contribution in (\ref{Jones.1order}) associated to the factor $\int_M A^{(0)} \wedge
d A^{(1)}$, then the term to compute is the following

\begin{multline}
\label{1.2pathexpanded} \frac{2ik}{\hbar N}\int \mathcal{D} A^{(0)}
\exp \left[\frac{i}{\hbar}k \int_{M} (A^{(0)} \wedge dA^{(0)} +
\frac{1}{k} A^{(0)} \wedge d \eta) \right] \left[  \int_M A^{(0)} \wedge d A^{(1)} \right] \\
= \frac{2k}{N} \int \prod_m d a^{(0)}_m \exp \left[ \frac{i}{\hbar}
k\left( \sum_m \lambda_m [(a^{(0)}_m)^2 + \frac{1}{k} a^{(0)}_m
\eta_m] \right)\right]
\\
\times \left[ \frac{3}{8} \frac{i}{\hbar} \Theta^{\kappa \lambda}
\sum_{i,j,\ell} \lambda_j \lambda_\ell a_i^0 a^0_j a^{(0)}_\ell
[C_{ij \ell}]_{\kappa \lambda} \right].
\end{multline}
The integral has the form in terms of $A_{ij}$ and $J_i$,  then we
get the expression


\begin{multline}
\label{final.2} \frac{2ik}{\hbar} \int \prod_m d a^{(0)}_m \exp
\left[ \frac{i}{\hbar} k\left( \sum_m \lambda_m \left[(a^{(0)}_m)^2
+ \frac{1}{k} a^{(0)}_m \eta_m \right] \right)\right] \cdot \left[
\frac{3}{8} \Theta^{\kappa \lambda} \sum_{i,j,\ell} \lambda_j
\lambda_\ell a_i^0 a^0_j a^{(0)}_\ell
 [C_{ij \ell}]_{\kappa \lambda} \right] \\
= \frac{3}{8}  \Theta^{\kappa \lambda} \sum_{i,j,\ell}  \lambda_j
\lambda_\ell [C_{ij \ell}]_{\kappa \lambda}  \exp \left(-\frac{i}{4
\hbar k} \int_M \eta \wedge d \eta \right) \\
\times \left[ - \frac{i}{4 \hbar k^2} \eta_{i} \eta_{j} \eta_{\ell}
+ \frac{1}{2 k} \left( \frac{\delta_{i \ell} \eta_j}{\lambda_i} +
\frac{\delta_{j\ell} \eta_i}{\lambda_j} + \frac{\delta_{i j}
\eta_\ell}{\lambda_j} \right) \right].
\end{multline}
Thus, the total contribution to the Jones-like polynomial up to
first order in $\Theta$ defined in (\ref{Jones.1order}) is obtained
from the superposition of equations (\ref{arbol}),
(\ref{1.2pathexpanded}) and (\ref{final.2}), yielding to the
expressions


\begin{equation}
J_C(\Theta) =  \exp \left(-\frac{i}{4 \hbar k} \int_M \eta \wedge d \eta
\right)  \left[1 +\frac{3}{16k}\Theta^{\kappa \lambda} \sum_{i,j}
\left( [C_{iji}]_{\kappa \lambda}  \lambda_j \eta_j +
[C_{ijj}]_{\kappa \lambda}  \lambda_j \eta_i \right) \right].
\label{finalJones}
\end{equation}
As we can see the zero-th order is the usual $U(1)$ ''Jones''
polynomial, where $\eta$ is the Poincar\'e dual of $C$ and the
following term is a polynomial over $\Theta$ related to the
noncommutativity up to first order.

\section{Noncommutative Aharonov-Bohm Effect}

This section is devoted to explore some physical applications of the
noncommutative Wilson loops and linking numbers, in particular we
consider the Aharonov-Bohm effect which is a very good arena to test
the physical ideas and extract visible effects. We are aware that
this subject is present in the literature, see for instance
\cite{chaichian1,gamboa,chaichian2,Falomir,li,dulat}. We shall see
that our results will be agree with their results. Aharonov-Bohm
effect consists of an electron beam through a double slit in
presence of a small impenetrable solenoid which has a non-vanishing
constant magnetic field inside (and therefore a non-vanishing vector
potential $A^{(0)}_\mu$). Outside the solenoid the magnetic field is
zero, but not the potential, thus an interference pattern is
observed due to the fact that the vector potential is non-vanishing.
The effect is measured as a phase factor in the wave function.

In the usual Aharonov-Bohm effect it is assumed that the wave
function is of the form $\Phi = \phi \exp(F)$. Under this ansatz one
finds the value of $F$ by means of the covariant derivative $D_j
\exp(F) = k_j \exp(F)$. In our case, we will assume that the
corresponding noncommutative function $\widehat{F}$ can be expanded
in terms a noncommutative expansion in terms of the noncommutative
parameter $\Theta$; given by the ansatz \cite{Falomir}
$\widehat{F}=F^{(0)} + F^{(1)} + F^{(2)} + \cdots$. We will also
determine $\widehat{F}$ using the covariant derivative defined in
the first section. Thus we have

\begin{equation}
D_j \star \exp_{\star}(\widehat{F}) = k_j \exp_{\star}(\widehat{F}),
\end{equation}
considering just the expansion up to second order, it can be written
up to second order as


\begin{multline}
\partial_j \left[ \exp(F^{(0)}) \left( 1 + F^{(1)} + F^{(2)} + \frac{1}{2} (F^{(1)})^2 \right) \right]
 - i(A_j^{(0)} + A_j^{(1)} + A_j^{(2)}) \exp(F^{(0)}) \\  \times \left( 1 + F^{(1)} + F^{(2)} + \frac{1}{2} (F^{(1)})^2 \right)
+  \frac{1}{2} \Theta^{\kappa \lambda} [\partial_\kappa(A_j^{(0)} +
A_j^{(1)})][\partial_\lambda(F^{(0)} + F^{(1)})] = k_j.
\end{multline}
Thus we obtain the following equations at each order

\begin{eqnarray}
\partial_j F^{(0)} - iA_j^{(0)} = k_j, \\
\partial_j F^{(1)} - i A_j^{(1)} + \frac{1}{2} \Theta^{\kappa \lambda} (\partial_\kappa A_j^{(0)})(\partial_\lambda F^{(0)}) = 0, \\
\partial_j F^{(2)} - i A_j^{(2)} + \frac{1}{2} \Theta^{\kappa \lambda} (\partial_\kappa A_j^{(0)})( \partial_\lambda F^{(1)}) + \frac{1}{2} \Theta^{\kappa \lambda} (\partial_\kappa A_j^{(1)})(\partial_\lambda F^{(0)}) = 0.
\end{eqnarray}
We can solve $F^{(0)}$ in terms of $A_j^{(0)}$, then we solve for
$F^{(1)}$ in terms of $A^{(1)}$ and $F^{(0)}$, and son on. In this
way we find at each order the $F^{(i)}$'s, explicitly

\begin{eqnarray}
F^{(0)} = k_j x^j + i \int_C A^{(0)}_j dx^j, \nonumber \\
F^{(1)} = i \int_C A_j^{(1)} dx^j - \frac{1}{2} \Theta^{\kappa
\lambda} \int_C
(\partial_\kappa A_j^{(0)})(\partial_\lambda F^{(0)}) dx^j, \nonumber \\
F^{(2)} = i \int_C A^{(2)}_j d x^j - \frac{1}{2} \Theta^{\kappa \lambda} \int_C (\partial_\kappa A_j^{(0)})(\partial_\lambda F^{(1)}) d x^j- \frac{1}{2} \Theta^{\kappa \lambda} \int_C (\partial_\kappa A_j^{(1)})(\partial_\lambda F^{(0)}).  \nonumber \\
\end{eqnarray}
Finally the expression for $\widehat{F}$ up to second order is given
by

\begin{multline}
\label{efe} \widehat{F} =  k_j x^j + i \int_C (A^{(0)}_j+ A^{(1)}_j+
A^{(2)}_j) dx^j - \frac{1}{2} \Theta^{\kappa \lambda}  \int_C \Big[
(\partial_\kappa A_j^{(0)})(\partial_\lambda F^{(0)}) \\ +
(\partial_\kappa A_j^{(0)})(\partial_\lambda F^{(1)}) +
(\partial_\kappa A_j^{(1)})(\partial_\lambda F^{(0)}) \Big] d x^j.
\end{multline}
From this equation we easily recognize the second term which
correspond to the second order expansion of the noncommutative
Wilson loop, and the following terms are the noncommutative
corrections to the holonomy.

In the usual Aharonov-Bohm effect the potential outside the solenoid
is given by
\begin{eqnarray}
A_1^{(0)} = -\frac{x_2}{x_1^2 + x_2^2}, \quad A_2^{(0)} =
\frac{x_1}{x_1^2 + x_2^2},
\end{eqnarray}
for reference the expansion up to first and second order of this
potential is

\begin{eqnarray}
A_1^{(1)} = \frac{1}{2} \Theta^{12} \frac{x_2}{(x_1^2 + x_2^2)^2}, \quad A_2^{(1)} = -\frac{1}{2} \Theta^{12} \frac{x_1}{(x_1^2 + x_2^2)^2} ,\\
A_1^{(2)} = -\frac{1}{4} (\Theta^{12})^2 \frac{2 x_2^3 + x_1^2
x_2}{(x_1^2 + x_2^2)^4}, \quad A_2^{(2)} = \frac{1}{2}
(\Theta^{12})^2 \frac{2 x_1^3 + x_1 x_2^2}{(x_1^2 + x_2^2)^4}.
\end{eqnarray}
Meanwhile inside the components of the solenoid are

\begin{eqnarray}
\label{inside} A_1^{(0)} = -\frac{B}{2}x_2, \quad A_2^{(0)} =
\frac{B}{2}{x_1},
\end{eqnarray}
where $B$ is the magnitude of a constant magnetic field. At first
and second order the potential is given by:

\begin{eqnarray}
\label{inside1}
A_1^{(1)} = - \frac{3}{8}B^2 \Theta^{12} {x_2}, \quad A_2^{(1)} = \frac{3}{8} B^2 \Theta^{12} x_1,\\
\label{inside2} A_1^{(2)} = -\frac{5}{16}B^3 (\Theta^{12})^2 x_2,
\quad A_2^{(2)} = \frac{5}{16} B^3(\Theta^{12})^2 x_1.
\end{eqnarray}
In analogy to the usual Aharonov-Bohm effect the phase difference is
modified proportional to the flux through the solenoid. Then
substituting equations $(\ref{inside}, \ref{inside1},\ref{inside2})$
in $(\ref{efe})$ we obtain the wave function's phase of the
non-commutative Aharonov-Bohm effect,
\begin{equation}
\widehat{F} = k_j x^j + i \pi B r^2 + i \frac{1}{2} \Theta^{12} \pi
B r^2 + \frac{5}{8}i(\Theta^{12})^2 \pi B r^2,
\end{equation}
where $\Phi = \pi r^2 B$ is the flux of the magnetic field trough
the solenoid (whose radius is $r$). Finally the correction to the
phase due the noncommutativity is
\begin{equation}
\exp(\widehat{F}) \approx \exp (\frac{ie}{\hbar} k_j x^j) \exp
\bigg[\frac{ie}{\hbar} \Phi + \frac{3ie}{4 \hbar} \Phi \Theta^{12}
+ \frac{5ie}{8 \hbar} \Phi (\Theta^{12})^2 \bigg]. \label{phaseAB}
\end{equation}
As we can see the first term in the imaginary exponential is the
standard holonomy (commutative), the second and third terms are
correction to the holonomy due the noncommutativity up to second
order given in terms of the usual flux and the noncommutative
parameter.


Now we proceed to show that the phase of the equation
(\ref{phaseAB}) is related to the non-commutative Landau levels
showed in Ref. \cite{GSH}. To check closely this affirmation let us
make some explicit computations. First of all we define the
non-commutative canonical momentum $\widehat{\Pi}_\mu$ as usual but
changing the usual gauge field $A_\mu$ by its noncommutative one
$\widehat{A}_\mu$, i.e.
\begin{equation}
\label{canonical} \widehat{\Pi}_{\mu} = p_\mu + e \widehat{A}_\mu =
p_\mu + e (A^{(0)}_\mu + A^{(1)}_\mu+A^{(2)}_\mu + \cdots).
\end{equation}
In addition, the consideration of quantum mechanical systems in a
non-commutative space leads to the following commutation relations
\begin{equation}
[p_\mu,p_\nu] = 0 , \quad [x_\mu,x_\nu] = 0 , \quad [p_\mu, x_\mu] =
- i \hbar \delta_{\mu \nu}.
\end{equation}
With the aid of previous relations and baring in mind the
expressions $(\ref{inside}-\ref{inside2})$ let us compute the
following commutator up to first order in $\Theta$, this is given by
\begin{equation}
\label{com-momentum} [\widehat{\Pi}_1, \widehat{\Pi}_2] = - i e
\hbar B(1 + \frac{3}{4}\Theta^{12} ).
\end{equation}
Let us construct the creation-annihilation operators as
\begin{equation}
a = \frac{\widehat{\Pi}_1 - i \widehat{\Pi}_2}{\sqrt{(2e \hbar B)(1 +
\frac{3}{4}\Theta^{12} )}}, \quad a^{\dagger} = \frac{\widehat{\Pi}_1 + i
\widehat{\Pi}_2}{\sqrt{(2e \hbar B)(1 + \frac{3}{4}\Theta^{12} )}},
\end{equation}
which satisfy the usual relation $[a,a^{\dagger}] = 1$. Now we will
consider the hamiltonian $H = \frac{1}{2 m} (\Pi^2_1 + \Pi^2_2)$ and
rewrite it in terms of $a$ and $a^{\dagger}$
\begin{equation}
H = \frac{eB \hbar}{2 m}(1 + \frac{3}{4} \Theta^{12}) (aa^{\dagger}+
a^{\dagger}a),
\end{equation}
where the factor $\omega=\frac{eB }{2 m}(1 + \frac{3}{4}
\Theta^{12})$ is the angular frequency. Using the normal ordering we
finally get
\begin{equation}
H = \hbar \omega (a^{\dagger}a + \frac{1}{2}).
\end{equation}
The examination of the phase in (\ref{phaseAB}) up to first order,
it clearly contains the frequency of the non-commutative Landau
levels.

To estimate the order of $\Theta$ we can compare with the results in
reference \cite{Falomir}, where they don't use the Seiberg-Witten
map. Similarly as Ref. \cite{Falomir}, it is possible to formulate
the problem of scattering charged particles in an effective radial
potential. The computation is exactly the same. Thus, at the first
order we obtain the same result than in \cite{Falomir} if we make
the following substitution: $\Theta$ by $-3 \Theta^{12}$. Then we
conclude that the non-commutative parameter is precisely of the same
order of magnitude as they estimated $\Theta^{12} \approx
[10\,Tev]^{-2}$.


\section{Final Remarks}

In this paper we proposed to use the gauge field provided by the Seiberg-Witten map to construct study noncommutative Wilson loops. After a brief account on noncommutative
Wilson loops, we study abelian Chern-Simons theory on a three
dimensional manifold. It was shown that the effect of
noncommutativity is the appearance of $6^n$ new knots at the $n$-th
order of the Seiberg-Witten expansion. These knots constitute
trivial homology cycles which are Poincar\'e dual to the high-order
Seiberg-Witten potentials of the expansion. Moreover the linking
number at $n$-th order of a standard 1-cycle with the multiple
Poincar\'e dual of the gauge fields is shown to be written as the
sum of the linking number of this 1-cycle with the multiple
Poincar\'e dual of the Seiberg-Witten gauge fields at this order
(\ref{newknots}). The generalization
to higher dimensions can be done straightforwardly.

Furthermore as a topological application of the noncommutative gauge
theories and Wilson loops in the abelian case; by using the path
integral formalism and the Chern-Simons theory we computed the first
order and non-vanishing correction due to the noncommutativity of
the abelian Jones-like polynomials (\ref{finalJones}). This term is
also of a topological nature and it represents a new noncommutative
topological effect of link invariants of three-manifolds.

Furthermore as a physical application we compute explicitly the abelian
Aharonov-Bohm effect in $\IR^3$ and calculate the wave function up
to second order in the noncommutativity parameter (\ref{phaseAB}).
It results in the usual wave function in terms of the
eigenvalue $k_j$ in the real exponential and in the imaginary part
appears the contribution of the usual flux (commutative) and a
second order contribution which is proportional to the square of the
flux. We discuss the relation of the
Aharonov-Bohm effect and the Landau levels in the non-commutative
context. These results are found to agree with those found in Refs.
\cite{chaichian1,gamboa,chaichian2,Falomir,li,dulat}. In particular,
the parameter $\Theta$ is constrained and it basically coincides
with that of Ref. \cite{Falomir}.

It could be interesting to explore some geometrical aspects we might
extend the linking number between knots in the three dimensional
Euclidean space in a noncommutative sense and explore the
different orders of the gauge potential and how they could give new information about linking numbers.

Moreover we would like to extend the present noncommutative ideas to
higher dimensional theories, trough a $BF$ theory since it is a
higher-dimensional generalization of Chern-Simons theory and   the
Wilson line will be interpreted in terms of linking numbers between
higher-dimensional objects \cite{HS,GCSS}.

Wilson loops for the spin connection are very important in some
theories of quantum gravity
\cite{rovelli,ashtekar,edwitten,witten-kodama}. It is worth to study
a noncommutative version of these models by using the noncommutative
Wilson loops described here. Some of this work is left for future calculations.

As a further step we are interested in the natural extension to the
non-abelian case, where we will deal with two expansions: the first
one focusing in the noncommutative parameter and the second one due
the non-abelianity of the Chern-Simons theory.  Also we will study
the physical implications using non-abelian Aharonov-Bohm effect
\cite{Horvathy}. For future work we leave the problem of studying
the generalization of noncommutative Aharonov-Bohm effect and its
associated Landau levels for this non-abelian case.

\vskip 2truecm
\centerline{\bf Acknowledgments} \vspace{.5cm}

The work of H. G-C. was partially supported by the CONACyT research
grant: 128761. The work of O.O.  is supported by a PROMEP, CONACyT  and UG grants. In addition the
work of R. S-S. was partially supported by a PROMEP and CONACyT
postdoctoral fellowship.

\vskip 1truecm

\end{document}